# BInGaN alloys nearly lattice-matched to GaN for high-power high-efficiency visible LEDs

Logan Williams and Emmanouil Kioupakis[*]

*Department of Materials Science and Engineering, University of Michigan, Ann Arbor, MI, USA 48109*

InGaN-based visible LEDs find commercial applications for solid-state lighting and displays, but lattice mismatch limits the thickness of InGaN quantum wells that can be grown on GaN with high crystalline quality. Since narrower wells operate at a higher carrier density for a given current density, they increase the fraction of carriers lost to Auger recombination and lower the efficiency. The incorporation of boron, a smaller group-III element, into InGaN alloys is a promising method to eliminate the lattice mismatch and realize high-power, high-efficiency visible LEDs with thick active regions. In this work we apply predictive calculations based on hybrid density functional theory to investigate the thermodynamic, structural, and electronic properties of BInGaN alloys. Our results show that BInGaN alloys with a B:In ratio of 2:3 are better lattice matched to GaN compared to InGaN and, for indium fractions less than 0.2, nearly lattice matched. Deviations from Vegard's law appear as bowing of the in-plane lattice constant with respect to composition. Our thermodynamics calculations demonstrate that the solubility of boron is higher in InGaN than in pure GaN. Varying the Ga mole fraction while keeping the B:In ratio constant enables the adjustment of the (direct) gap in the 1.75-3.39 eV range, which covers the entire visible spectrum. Holes are strongly localized in non-bonded N 2p states caused by local bond planarization near boron atoms. Our results indicate that BInGaN alloys are promising for fabricating nitride heterostructures with thick active regions for high-power, high-efficiency LEDs.

InGaN light-emitting diodes (LEDs) with an electricity-to-light conversion efficiency of ~39% provide large efficiency gains and cost reductions compared to incandescent and fluorescent light sources.[1] However, InGaN LEDs suffer from decreasing internal quantum efficiency at high currents (efficiency droop), particularly at longer wavelengths (green gap). The cause of the droop has been extensively studied, with Auger recombination identified as a major loss mechanism.[2] For equal electron and hole densities, the Auger recombination rate is equal to the carrier density cubed times a material-dependent Auger coefficient $C$. Since the Auger coefficient is an intrinsic property of InGaN that does not depend strongly on composition, temperature, or strain,[3] the carrier density at a given current density must be lowered to reduce the Auger losses.

Increasing the active-region volume is a straightforward approach to reduce the carrier density and hence the Auger loss, yet growth challenges limit its practicality. Devices using a single thick InGaN layer exhibit higher high-power efficiency than thinner quantum wells both for polar[4] and for semipolar[5] growth orientations. The thickness of InGaN active layers is however limited by the lattice mismatch with the underlying GaN layers, and the subsequent appearance of performance-degrading dislocations. On the other hand, the efficiency of multiple-quantum-well (MQW) structures is lower than a single InGaN layer of the same total active thickness[5] since hole injection is poor and only the first few QWs near the p-layer in a MQW structure emit light.[6,7]

The co-alloying of InGaN with wurtzite boron nitride (w-BN) can produce BInGaN alloys lattice-matched to GaN with gaps spanning the visible range (Fig. 1). Co-alloying has been demonstrated in, e.g., GaAsPBi[8] and GaAsNBi,[9] in which the co-incorporation of P (N) and Bi atoms yields alloys lattice-matched to GaAs with a reduced band gap. Under ambient conditions, BN adopts the hexagonal layered structure that can be exfoliated to form 2D materials. Wurtzite BN is a high-pressure polytype with the same crystal structure as InGaN. For low BN content, alloys of BN with InGaN are expected to also adopt the wurtzite structure to minimize dangling-bond formation. Moreover, although w-BN is an indirect-gap semiconductor, the gap of BInGaN is determined by the lower gap of the InGaN component, which is direct. The in-plane lattice constant of BInGaN is approximated, to first order, by Vegard's law. Based on the experimental in-plane lattice constants of GaN (3.181 Å), InN (3.538 Å) and w-BN (2.536 Å),[10–12] the optimal B mole fraction to lattice-match $B_yIn_{1-y}N$ to GaN is y=0.356≈0.4. Therefore, an approximate ratio of B:In≈2:3 is expected to yield BInGaN alloys nearly-lattice-matched to GaN. Hence, varying the Ga mole fraction while keeping the B:In ratio constant is a promising method to tune the gap of BInGaN while keeping the in-plane lattice constant nearly matched to GaN.

Previous experimental and theoretical studies have explored the structural and electronic properties of B-containing GaN, InN, and InGaN. Ougazzaden et al. grew BGaN thin films with up to 3.6% boron.[13] They reported lower gaps than GaN and a large bowing parameter.[14] Kadys et al. used metal-organic chemical vapor deposition (MOCVD) to grow up to 2.9, 4.3, and 5.5% boron BGaN on GaN, AlN, and SiC substrates respectively.[15] Cramer et al. reported high-crystal-quality BGaN with up to 3% boron grown with plasma-assisted molecular beam epitaxy and observed statistically random atomic distributions.[16] Gunning et al. grew up to 7.4% boron BGaN on AlN using metalorganic vapor phase epitaxy (MOVPE) at low temperatures and pressure (750-900 °C and 20 Torr), but reported severe structural degradation and a shift to a twinned cubic structure at higher boron concentrations.[17] Approximately equimolar nanocolumnar BInN has also been synthesized with a reported gap of 3.75 eV.[18] Quaternary BInGaN has been grown on GaN/sapphire and ZnO-buffered Si substrates with MOVPE by Gautier, Ougazzaden, et al.[19,20] They reported smaller lattice constants and gaps than GaN for up to 2% B and up to 14% In content.[19,20] The extracted bowing parameters were applied to predict the gap for a broader composition range. However, their InGaN bowing parameter is larger than subsequent predictive calculations,[21] while the BN gap was set to the indirect Γ-K value instead of the larger direct Γ-Γ one.[13,14,22] McLaurin also reported the growth of BInGaN.[23] Theoretically, Park and Ahn examined BInGaN/GaN quantum wells with effective mass theory. They reported a lower mismatch to GaN than InGaN, and a reduction of the polarization fields.[24] Assali et al. examined ordered BInGaN in the metastable zinc blende phase with density functional theory (DFT). They found a near-lattice-match to GaN for a composition with the same 2:3 B:In ratio as our estimate, and a small increase of the gap (0.1-0.3 eV) compared to InGaN of the same In content.[25] However, they did not examine the thermodynamically stable wurtzite phase, which has a larger gap than zinc blende,[26] and did not account for disorder. Overall, the properties of GaN-lattice-matched disordered wurtzite BInGaN alloys over their full composition range and their potential for reducing the LED droop problems remain unexplored.

In this work, we explore the thermodynamic, structural, and electronic properties of statistically random quaternary wurtzite BInGaN alloys with hybrid-functional DFT. BInGaN alloys with a B:In ratio of 2:3 are better lattice mismatch to GaN than InGaN. Co-alloying with In lowers the enthalpy of mixing and facilitates higher B incorporation. Our results show that BInGaN alloys can be designed nearly-lattice-matched to GaN with a direct band gap adjustable over the entire visible range, and are therefore promising active-layer materials to overcome the efficiency-droop and green-gap problems of nitride LEDs.

We performed DFT calculations based on the projector augmented wave (PAW) method[27,28] using the Vienna Ab initio Simulation Package (VASP).[29–32] The GW-compatible pseudopotentials including 3, 13, 13, and 5 valence electrons were employed for B, In, Ga, and N, respectively, with a 600 eV plane-wave cutoff. Structural relaxations were performed using the optB86b-vdW functional[33] and a Γ-centered Wisesa-McGill-Mueller Brillouin-zone grid with a minimum period distance of 21.48 Å.[34] Forces on atoms were relaxed to 1 meV/Å. Band-gap calculations were performed with the Heyd–Scuseria–Ernzerhof (HSE06) functional.[35,36] Random alloys were modeled using Special Quasi-random Structures (SQS) generated with the Alloy Theoretic Automated Toolkit[37] and a 3×3×2 wurtzite supercell. Cations were arranged to approximate the pair-correlation functions of random alloys up to 5.125 Å. Five SQSs were generated at each composition and relaxed to obtain structural parameters. The SQSs that most closely match random pair-correlation functions at each composition were used for electronic and thermodynamic calculations (Fig. S1). The projected density of states (pDOS) was calculated using the optB86b-vdW functional and a Γ-centered 8×8×8

Brillouin-zone grid, and the gap was rigidly shifted to the HSE06 value.

Our thermodynamic analysis reveals that the solubility of B into InGaN is higher than into GaN. To calculate the transition temperature between the solid-solution and the miscibility-gap regimes as a function of composition, $T(x) = \Delta H(x)/S$, we evaluated the enthalpy of mixing as a function of alloy composition, $\Delta H(x)$, by subtracting the total energy of ternary and quaternary alloys from the linear combination of the binaries. The entropy was evaluated using the regular solution model, $S = -k_B \sum_{i=1}^{N} x_i \ln x_i$ ($x_i$ is the mole fraction for each of the $N$ alloy ingredients, and $k_B$ is Boltzmann's constant).. The transition temperatures (Fig. 2) for $B_xIn_{1.5x}Ga_{1-2.5x}N$, $B_xGa_{1-x}N$, and $In_{1.5x}Ga_{1-1.5x}N$ are well above typical growth temperatures for x > 0.05, which is expected since nitrides are typically grown with epitaxial techniques (e.g., MOCVD or molecular beam epitaxy) that take place far from thermodynamic equilibrium. However, the relative equilibrium phase boundaries can be combined with experimental literature data to predict the limits of boron incorporation into InGaN. For boron mole fractions less than 0.2, the temperature needed to dissolve B into InGaN is approximately 2/3 of the temperature needed to dissolve it into GaN. The increased solubility of B into InGaN is due partly to the increased configurational entropy in the quaternary alloy, and partly to the partial cancellation of local stress by the opposite size mismatch between the smaller B and the larger In atoms. We therefore estimate that 1.5× as much boron can be incorporated into InGaN compared to GaN. Since high-quality, single-phase $B_xGa_{1-x}N$ alloys with boron concentration up to x=0.03 have already been demonstrated,[16] we anticipate that up to ~4.5% boron incorporation should be possible in BInGaN with existing growth approaches.

Our calculations verify that BInGaN with a 2:3 B:In ratio is better lattice-matched to GaN substrates than InGaN and can therefore be grown to larger thicknesses. Figure 3 shows the calculated configurationally averaged $a$ and $c$ lattice constants of $B_xIn_{1.5x}Ga_{1-2.5x}N$ relative to GaN. The trend for $a$ is not described well by a straight line, and therefore deviates from Vegard's law, agreeing with previous calculations for BGaN.[38] Instead, it follows a bowing relationship, $a_{BInGaN}(x)/a_{GaN} = a_1 x + a_2(1-x) - bx(1-x)$, where $a_1 = 1.034$, $a_2 = 1$, and $b = 0.052$. The $a$ lattice-constant mismatch of BInGaN is smaller than +/–0.25% for indium fractions under 0.2 (Fig. 3(a)), which facilitates growth on the $c$-plane of GaN. Due to the bowing, the mismatch of $B_{0.045}In_{0.0675}Ga_{0.8875}N$ to GaN is predicted to be only – 0.1%. The $c$ lattice constant of BInGaN (Fig. 3(b)) remains near the GaN value.

Our band-gap calculations demonstrate that $B_xIn_{1.5x}Ga_{1-2.5x}N$ has a tunable direct gap that spans the entire visible range. In contrast to BGaN, which transitions from direct to indirect gap for increasing B content,[39] the gap of BInGaN remains direct throughout the entire explored composition range. Figure 4 shows the calculated gap of $B_xIn_{1.5x}Ga_{1-2.5x}N$, $B_xGa_{1-x}N$, and $In_{1.5x}Ga_{1-1.5x}N$ as a function of composition. For the 2:3 B:In ratio, the gap ranges from 3.14 to 1.50 eV for decreasing Ga content. Increasing our calculated values by 0.25 eV to match the room-temperature gap of GaN (3.39 eV)[40] brings our BInGaN gap estimate to the 3.39–1.75 eV range, spanning the entire visible spectrum. The gap of $B_xIn_{1.5x}Ga_{1-2.5x}N$ is also approximately equal to that of $In_{1.5x}Ga_{1-1.5x}N$ for the same In mole fraction. To explore this behavior, we determined the orbital character of the conduction and valence band edges. The pDOS of $B_{0.278}In_{0.417}Ga_{0.306}N$, calculated with the optB86b-vdW functional, (Fig. 5; comparison to the HSE pDOS in Fig. S2) shows that the edge states consist primarily of N and In orbitals, while B states lie higher in the conduction band. Similar to InGaN, the valence-band edge consists primarily of localized N 2$p$ states, except in BInGaN the states reside near planarized B atoms (Fig. 5 inset). The hole localization energy ranges from 0.1-0.4 eV as the boron content ranges from 15-40%, the maximum value occurring for $B_{0.278}In_{0.417}Ga_{0.306}N$ (Fig. 6). Localized states were not observed for boron mole fractions lower than 15%. The conduction-band edge is primarily formed by N 2$p$, N 2$s$, and In 5$s$ states. Our calculations therefore show that partial substitution of B for Ga in InGaN has only minor effects on the gap and edge states.

The addition of boron into InGaN active layers is a promising method of overcoming the LED droop and green-gap problems. The better lattice match with GaN and the resulting reduced strain allows the growth of thicker active layers, thus decreasing the carrier concentration for a given current density and hence the fraction of carriers that recombine via Auger. While the spatial separation of carriers by the polarization fields in polar wells is amplified for thicker layers at low carrier concentrations, the polarization fields are screened by free carriers under high-power operation and flat-band conditions prevail.[2] Our estimate for the Debye screening length $L_D = \sqrt{\frac{k_B T}{4\pi n e^2}}$ is approximately 0.5 nm for typical carrier densities ($n \cong 5\times10^{18}$ cm$^{-3}$) of the internal quantum efficiency maximum.[2] Therefore, for sufficiently thick wells (e.g., thicker than 10 nm) and high carrier densities the polarization fields are completely screened by free carriers.

One challenge regarding the growth of BInGaN is the different temperatures needed for the ingredient materials. GaN is typically grown at high temperature to achieve higher crystalline quality, but InGaN and BGaN are typically grown at lower temperatures to facilitate In and B incorporation.[15,17] However, the lower enthalpy of mixing B into quaternary BInGaN than in ternary BGaN may facilitate the growth of BInGaN in a wider temperature window. E.g., Gautier, Ougazzaden, et al. successfully grew BInGaN with up to 2% boron and 14% indium using MOVPE at 730 °C.[19,20] Another potential challenge is the appearance of secondary phases during growth. Gunning et al. found that BGaN creates a twinned cubic structure at their selected growth conditions.[17] Similar cubic secondary phase inclusions may also form during BInGaN growth and deteriorate the structural quality. Hence, the thermodynamics of both hexagonal and cubic phases of BInGaN need to be further investigated both experimentally and theoretically to facilitate the development of device-quality materials.

In conclusion, we examine the effects of co-alloying boron and indium into GaN with first-principles calculations. Alloying 2B:3In into GaN creates alloys with gaps similar to InGaN of the same indium concentration, while reducing lattice mismatch and nearly eliminating it for alloys with In mole fractions less than 0.2 (i.e., band gaps larger than 2.75 eV). Our thermodynamics analysis reveals that B is more easily incorporated into InGaN than into pure GaN. Our results point to BInGaN alloys as promising materials to fabricate thicker active regions than InGaN for higher-efficiency high-power visible LEDs.

See supplementary material for technical details on the band-gap variability between SQSs and on the comparison of the optB86b-vdW and the HSE06 pDOS.

This work was supported by the Designing Materials to Revolutionize and Engineer our Future (DMREF) Program under Award No. 1534221, funded by the National Science Foundation. This research used resources of the National Energy Research Scientific Computing (NERSC) Center, a DOE Office of Science User Facility supported under Contract No. DE-AC02-05CH11231.

* kioup@umich.edu

FIGURE CAPTIONS

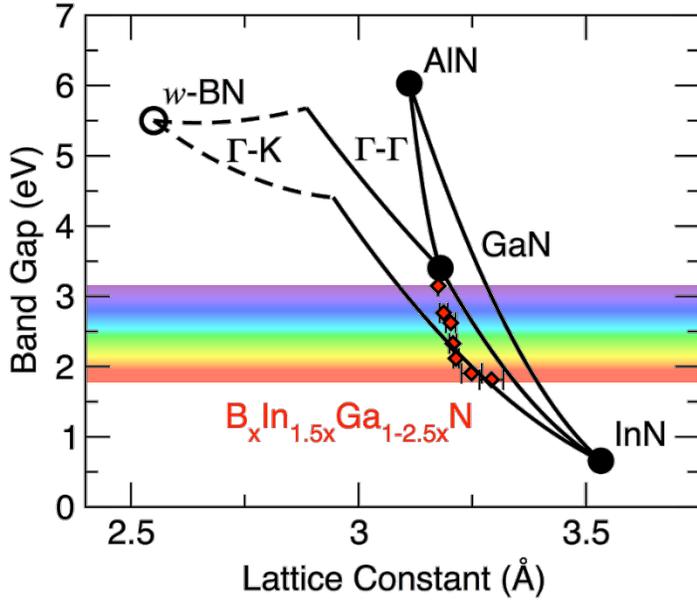

FIG. 1. Band gaps vs. in-plane lattice constant (*a*) for wurtzite group-III nitrides. The $B_xIn_{1.5x}Ga_{1-2.5x}N$ alloys investigated in this work (red diamonds) maintain approximate lattice match to GaN while their gaps span the entire visible range. The error bars show the uncertainty in the lattice constant (see Figure 3). Closed circles and full lines represent direct gap materials. Open circles and dashed lines represent indirect gap materials.

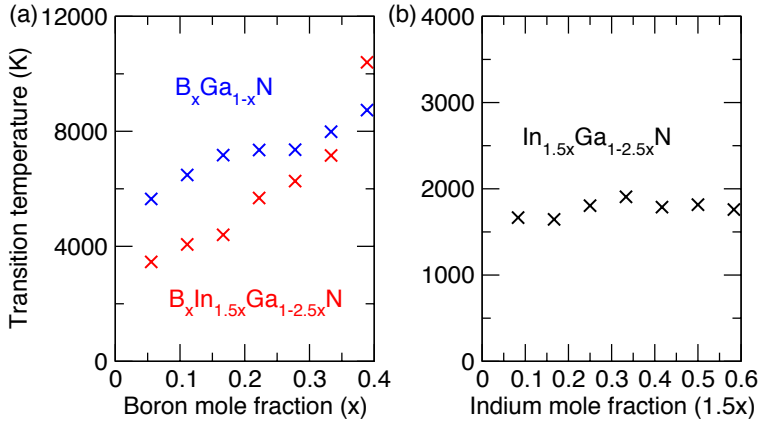

FIG. 2. Calculated transition temperature for the thermodynamic equilibrium random mixing of group III nitride alloys. (a) $B_xIn_{1.5x}Ga_{1-2.5x}N$ and $B_xGa_{1-x}N$ as a function of Boron mole fraction. (b) $In_{1.5x}Ga_{1-1.5x}N$ as a function of Indium mole fraction. The transition temperatures of $B_xIn_{1.5x}Ga_{1-2.5x}N$ are approximately 2/3 that of $B_xGa_{1-x}N$ of equal B mole fraction at low boron concentrations (<0.2), indicating that boron is more easily incorporated into InGaN than into GaN.

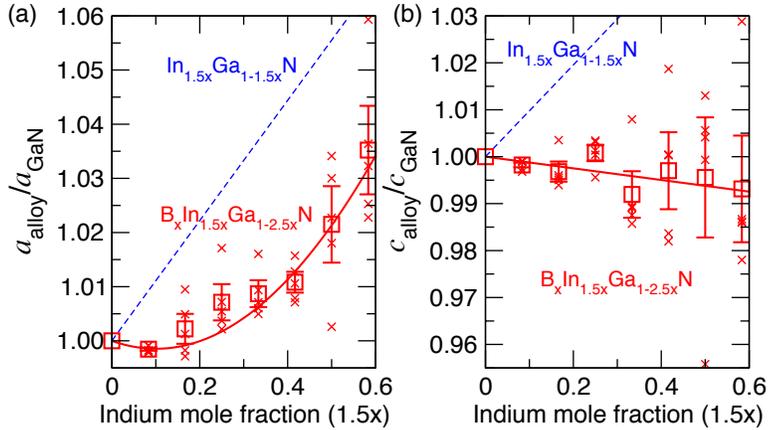

FIG. 3. The calculated lattice constants of InGaN and BInGaN alloys relative to GaN along (a) the *a* direction and (b) the *c* direction. Both the values for each configuration (crosses) and the configurational average (squares) are displayed. The *a* lattice constant data is fit to a bowing equation and has a bowing parameter of 0.052. The error bars show the statistical uncertainty for each configurational average, calculated as half of the range divided by the square root of the number of samples.. The mismatch of $B_xIn_{1.5x}Ga_{1-2.5x}N$ to GaN along the *a* axis is significantly reduced compared to an equivalent $In_{1.5x}Ga_{1-1.5x}N$ alloy, especially at lower boron and indium concentrations, while BInGaN is approximately lattice matched to GaN along the *c* direction.

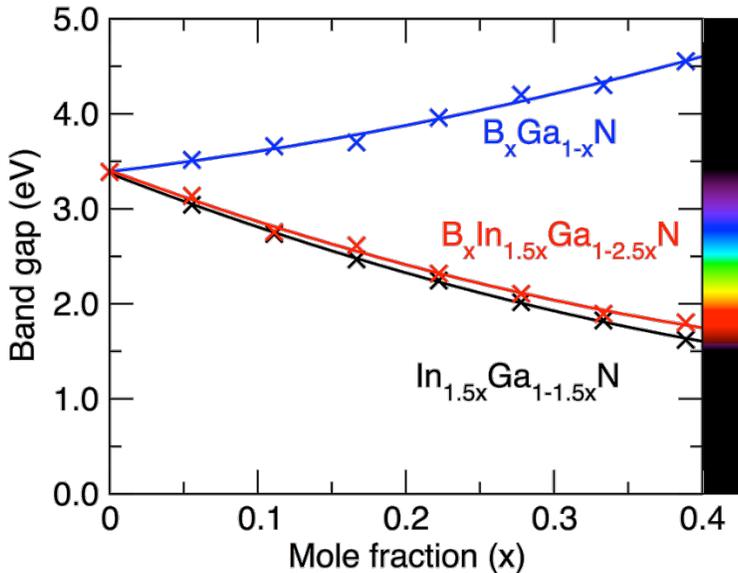

FIG. 4. The calculated band gaps of $B_xIn_{1.5x}Ga_{1-2.5x}N$, $In_{1.5x}Ga_{1-1.5x}N$, and $B_xGa_{1-x}N$ as a function of mole fraction x. The calculated gap values have been increased by 0.25 eV to match the experimental gap of GaN at room temperature (3.39 eV).[40] The band gap of BInGaN alloys spans the entire visible range. The gap of BInGaN has approximately the same value as an InGaN alloy of the same indium mole fraction, indicating that boron incorporation has negligible effect on the gap of InGaN. Figure S1 shows the variability in the DFT band gap for all simulation cells used in the structural calculations.

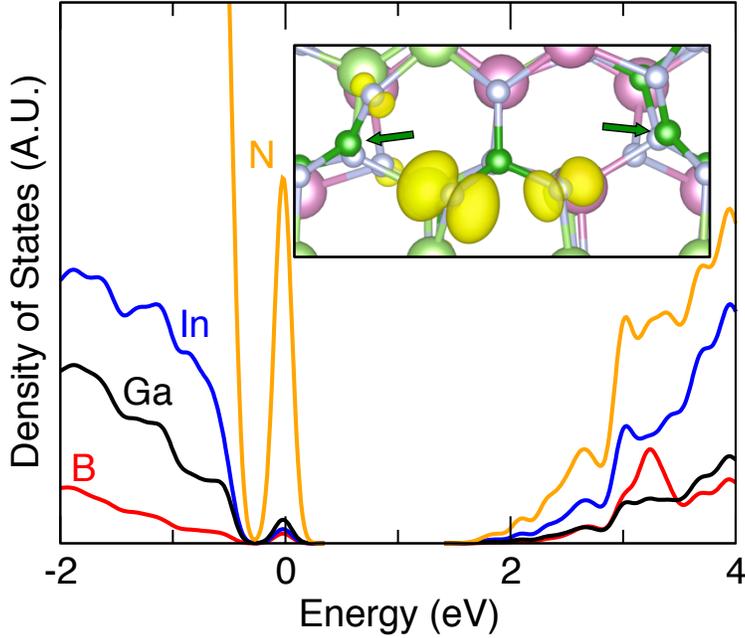

FIG. 5. Projected Density of States (pDOS) for a $B_{10}In_{15}Ga_{11}N_{36}$ solid solution. Similar to InGaN, the conduction-band edge is primarily composed of In and N states. The valence band displays a localized band of N 2p character caused by local planarization of B-N bonds near boron atoms. The electron density of the localized state and the planarized B atoms are visualized in the inset.. The pDOS calculated with HSE06 and optB86b-vdW are qualitatively similar, the only major quantitative difference being the band-gap value (Fig. S2).

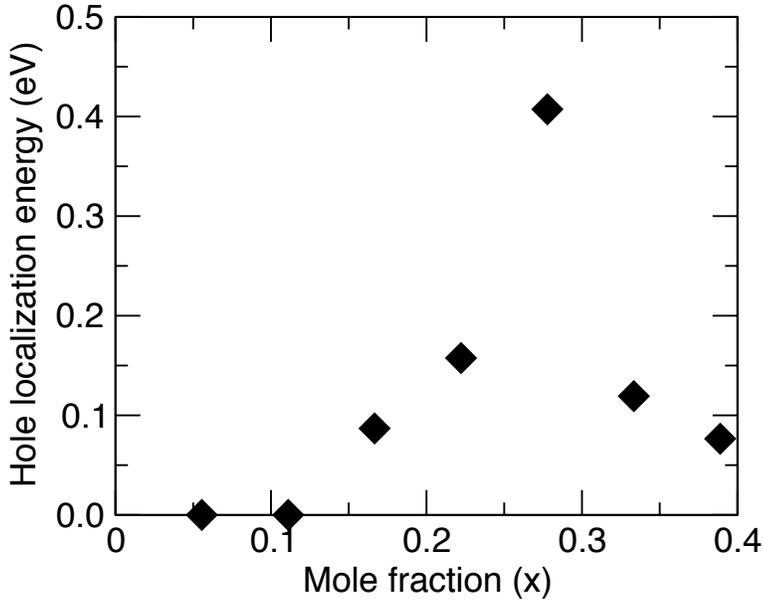

FIG. 6. Hole localization energy for $B_xIn_{1.5x}Ga_{1-2.5x}N$ calculated using HSE06. No localized states are seen at boron mole factions less than ~0.1. The hole localization energy is maximum near the $B_{0.278}In_{0.417}Ga_{0.306}N$ composition.

# BInGaN alloys nearly-lattice-matched to GaN for high-power high-efficiency visible LEDs

Logan Williams and Emmanouil Kioupakis[*]

*Department of Materials Science and Engineering, University of Michigan, Ann Arbor, MI, USA 48109*

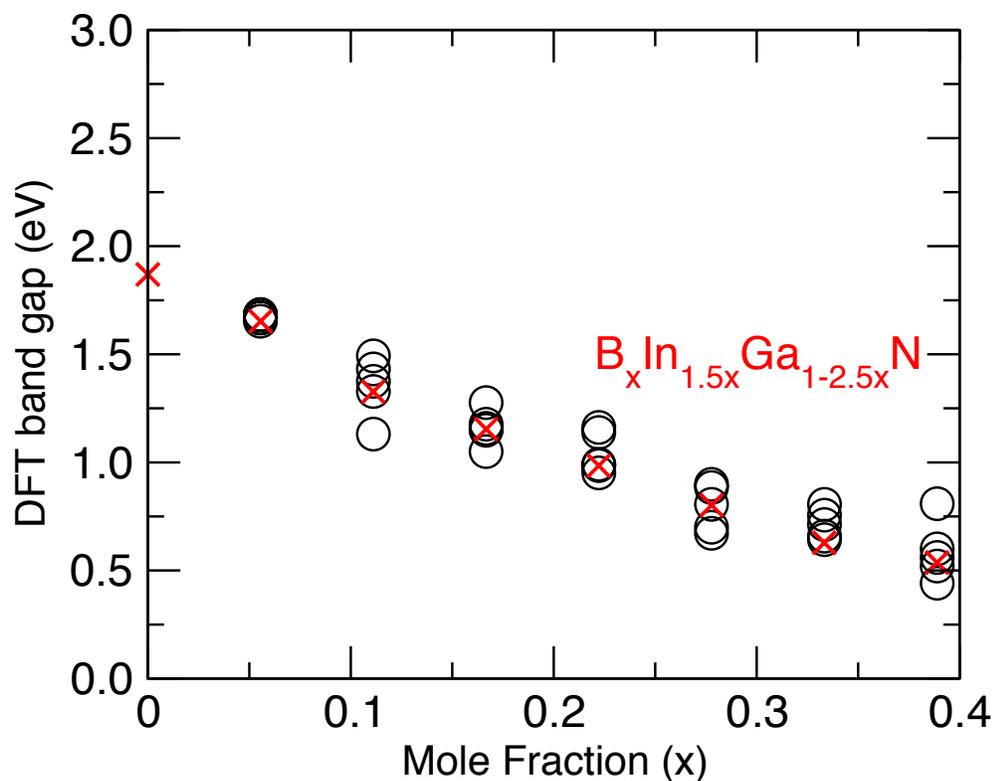

FIG. S1. Band gap of $B_xIn_{1.5x}Ga_{1-2.5x}N$ vs. boron mole fraction calculated with the optB86b-vdW functional for all cells used in the structural-analysis calculations. The band gap for each configuration is shown using black circles, while the red crosses denote the cells that most closely match the random pair-correlation functions at each composition that were subsequently used for the HSE06 electronic-structure calculations (Figures 4 and 6).

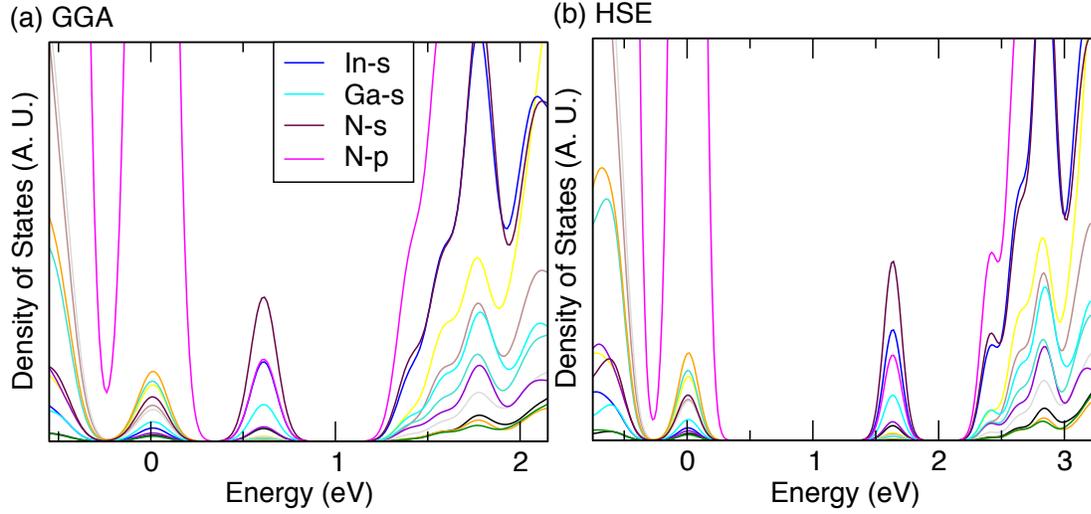

FIG. S2. Partial Density of States (pDOS) of a $B_{10}In_{15}Ga_{11}N_{36}$ solid solution calculated (a) with the optB86b-vdW functional and (b) with the HSE06 functional, using a 2x2x2 Γ-centered Brillouin-zone sampling grid. Energies are referenced with respect to the highest occupied valence band state. Both functionals display qualitatively similar pDOS, with the only significant quantitative difference being the increased value of the band gap with HSE. The gap observed in the conduction band is an artifact caused by under-sampling of the Brillouin zone. For this reason, the optB86b-vdW functional with a rigid shift applied to correct for the band-gap underestimation was used to converge the pDOS in Figure 5 as a function of Brillouin-zone sampling (with an 8x8x8 Γ-centered grid). The hole localization energy with optB86b-vdW is 0.354 eV, while HSE with a 25% Hartree-Fock mixing parameter increases the localization energy by ~15% to 0.407 eV. Using a mixing parameter of 29.63%, which causes the calculated band gap of GaN to match experiment, has a negligible effect (~4 meV change) on the localization energy.